\documentclass[12pt]{article}
\usepackage[utf8]{inputenc}
\usepackage{graphicx}
\usepackage{pgfplots}
\usepackage{amsmath}
\usepackage{amsfonts}
\usepackage{amssymb}
\usepackage{biblatex} 
\usepackage{csvsimple}
\usepackage{float}
\title{Computing Sensitivities in Reaction Networks using Finite Difference Methods}
\author{Evan Yip and Herbert Sauro}
\date{May 2021}

\begin{document}

\maketitle

\section{Abstract}

In this article, we investigate various numerical methods for computing scaled or logarithmic sensitivities of the form $\partial \ln y/\partial \ln x$. The methods tested include One Point, Two Point, Five Point, and the Richardson Extrapolation. The different methods were applied to a variety of mathematical functions as well as a reaction network model. The algorithms were validated by comparing results with known analytical solutions for functions and using the Reder method for computing the sensitivities in reaction networks via the Tellurium package. For evaluation, two aspects were looked at, accuracy and time taken to compute the sensitivities. Of the four methods, Richardson's extrapolation was by far the most accurate but also the slowest in terms of performance. For fast, reasonably accurate estimates, we recommend the two-point method. For most other cases where the derivatives are changing rapidly, the five-point method is a good choice, although it is three times slower than the two-point method. For ultimate accuracy which would apply particularly to very fast changing derivatives the Richardson method is without doubt the best, but it is seven-times slower than the two point method. We do not recommend the one-point method in any circumstance. The Python software that was used in the study with documentation is available at: \url{https://github.com/evanyfyip/SensitivityAnalysis}.

\section{Introduction}

In this study we will look at a number of different approaches to computing steady-state sensitivities in reaction networks~\cite{sauro2014Modeling,somogyi2015libroadrunner}. We are especially interested in the reaction networks found in biological systems, where they are often catalyzed by enzymes. In these cases, computing sensitivities is an effective way to gauge the role of the different steps in a regulated biochemical pathway.

Reaction networks are composed of one of more species nodes, $x_i$, connected via chemical reaction steps. For example, a simple chain of reaction steps is shown below:
$$ x_1 \stackrel{v_1}{\longrightarrow} x_2 \stackrel{v_2}{\longrightarrow} x_3 \stackrel{v_3}{\longrightarrow} x_4 \stackrel{v_4}{\longrightarrow} \dots \longrightarrow x_n $$
where $v_i$ is the reaction rate for the $i^{th}$ reaction. Such systems are often described by a set of differential equations (ODE), where each ODE describes the rate of change of a particular species. According to the conservation of mass, the rate of change of a given species must be the difference between the input rate and output rate. For example, in the above scheme, the rate of change of $x_2$ will be given by:
$$ \frac{dx_2}{dt} = v_1 - v_2 $$
Such equation can be devised for every species in the network.

In order to sustain a steady-state the sources and sinks must be clamped, that is, held constant. In the above scheme, $x_1$ and $x_n$ will be clamped. To be more specific, consider the simplest reaction network of two reactions:
$$ X_o \stackrel{v_1}{\longrightarrow} x_1 \stackrel{v_2}{\longrightarrow} X_1 $$
We will assume $X_o$ and $X_1$ are clamped and do not change in time. This means this system~\cite{sauro2014Modeling}:
\begin{eqnarray*}
 v_1 &= k_1 X_o - k_2 x_1 \\
 v_2 &= k_3 x_1 - k_4 X_1
\end{eqnarray*}
For simplicity, if we assume that the concentration of $X_1 = 0$, then the differential equation for $x_1$ is given by: 
$$ \frac{dx_1}{dt} = v_1 - v_2 = k_1 X_o - k_2 x_1 - k_3 x_1 $$
At steady-state $dx_1/dt = 0$. Setting the differential equation to zero gives:
$$ k_1 X_o - k_2 x_1 - k_3 x_1  = 0 $$
Solving for $x_1$ yields:
$$ x_1 = \frac{k_1 X_o}{k_2 + k_3} $$
We conclude from this that the steady-state is a function of all the parameters in the system. Moreover we can also derive the steady-state flux, $J$, through the system by substituting the steady-state value for $x_1$ into $v_1$. This gives:
$$ J = \frac{k_1 k_3 X_o}{k_2 + k_3} $$
Note that if the first reaction $v_1$ is irreversible then the steady-state flux reduces to $J = k_1 X_o$

Of interest is how the various constants influence the steady-state concentrations and fluxes. For example, if the reactions are enzyme catalyzed, it would be useful to know how much the activity of the various enzymes contribute to the steady-state flux and species concentrations. There are different ways to measure this influence, one way is to compute the unscaled derivatives:
$$ \frac{dJ}{dp}\ \mbox{ and }\ \frac{dx_i}{dp} $$
where $p$ is some parameter that we are interested in. An alternative formulation is to consider the scaled derivatives:
$$ \frac{dJ}{dp} \frac{p}{J}\ \mbox{ and }\ \frac{dx_i}{dp} \frac{p}{x_i} $$
These can be more useful because they are unit-less and also represent a ratio of fold changes which tends to be easier to measure experimentally. In the literature, these scaled sensitivities are referred to as control coefficients~\cite{kacser1995control,sauro2018systems}, where $p$ is the enzyme activity, or response coefficients when $p$ is for example an inhibitor such as a therapeutic drug. 
\begin{eqnarray*}
C^J_{e_i} &= \dfrac{dJ}{de_i} \dfrac{e_i}{J} \\[4pt]
C^{x_j}_{e_i} &= \dfrac{dx_j}{de_i} \dfrac{e_i}{x_j} \\
\end{eqnarray*}

In the simple network example above, we derived, with relative ease, the steady-state solution to the differential equation. However for large networks and especially those where the reaction rates $v_i$ are governed by nonlinear functions, it is not possible to derive closed solutions for the steady-state flux and steady-state. Instead, we must revert to numerical methods. In the remainder of the paper we will discuss various numerical methods that can be used to compute the control coefficients. Of particular interest are their accuracy and speed of computation.

\section{Methods}

There are various numerical methods for estimating the derivative of a function~\cite{chapra2011numerical}. The choice of which method to use depends on how accurate the estimate should be and how fast does the computation need to proceed. The simplest method for computing a derivative of a function $f(x)$ is the one-point method. 

\subsection*{One-Point Method}

All numerical methods for estimating derivatives are by their nature approximate.  The one-point method~\cite{chapra2011numerical}, also know as the Newton's difference quotient or forward-different method, is the simplest and is given by:
$$ \frac{df(x)}{dx} \approx \frac{f(x + h) - f(x)}{h} $$
The estimate relies on the step-size, $h$. A step-size that is too large, will incur a significant error in the estimate for the derivative especially when the function $f(x)$ is non-linear.  It's possible to estimate this error by expanding the function, $f(x)$ in a Taylor series:
\begin{equation}
f(x+h) = f(x) + h \frac{df(x)}{dx} + \mbox{ higher terms}
\label{eqn:1}
\end{equation}
The error in our estimate comes from neglecting the higher terms. If we assume this error is dominated by the first term in the higher terms (i.e the second derivative), then we can show that the error is given approximately by:
$$ \frac{h}{2} \frac{df(x)^2}{d^2t} $$
This tells us that the higher the value of $h$ the greater the error. It also indicates that the error is a function of the second derivative, that is the curvature. For a simple equation such as $y = x^2$, the curvature is a constant of value 2, so that the error term reduces to $h$. For $y = x^2$ the error in our estimate will therefore equal the step-size. 

Rearranging equation~\eqref{eqn:1} yields:
$$ \frac{df(x)}{dx}  = \frac{f(x + h) - f(x)}{h} - \mbox {higher terms} $$

In the limit, as the step size, $h$, tends to zero the approximation becomes exact:
$$ \frac{df(x)}{dx} = \lim_{h \to 0} \frac{f(x + h) - f(x)}{h}. $$

\subsection*{Two-Point Methods}

The one-point method is not recommended for estimating derivatives due to errors resulting from the higher terms. Instead a much more accurate estimate can be obtained by computing the value of the function on both sides of the center point. This is known as the two-point or central difference quotient~\cite{chapra2011numerical}. The formula for this is given by:
$$ \frac{df(x)}{dx}  \approx \frac{f(x + h) - f(x - h)}{2h} $$
In the one-point scheme the error was proportional to the size of the step, $h$. In the two-point case the error is proportional to the square of the step size $h$. One way to think of the two-point scheme is as a combination of a forward-difference and backward-difference method. When combined, the first-order terms cancel leaving the second order term intact. Hence the error become proportional to the square, or $h^2$. This can be shown as follows. First, we write out the forward and backward schemes as a Taylor series:
$$
\begin{aligned}
& f(x + h) \approx f(x) + h \cdot f^{\prime}(x)+(h)^{2} \frac{f^{\prime \prime}(x)}{2}+(h)^{3} \frac{f^{\prime \prime \prime}(x)}{6} \\[8pt]
& f(x - h) \approx f(x) - h \cdot f^{\prime}(x)+(h)^{2} \frac{f^{\prime \prime}(x)}{2}-(h)^{3} \frac{f^{\prime \prime \prime}(x)}{6} \\
\end{aligned}
$$
We then take the difference between them by computing $ f(x + h) - f(x - h)$ from which we obtain:

$$
\begin{aligned}
f(x + h)-f(x - h) \approx 2 h \cdot f^{\prime}(x)+(h)^{3} \frac{f^{\prime \prime \prime}(x)}{3} \\
\end{aligned}
$$
Rearranging this by dividing both sides by $2 h$ yields:
$$
\begin{aligned}
\frac{f(x + h) - f(x - h)}{2 h} \approx f^{\prime}(x) + (h)^{2} \frac{f^{\prime \prime \prime}(x)}{6}
\end{aligned}
$$
or
$$ f^{\prime}(x) \approx \frac{f(x + h) - f(x - h)}{2 h} -  (h)^{2} \frac{f^{\prime \prime \prime}(x)}{6}$$
We can see from this result that the error is proportional to $h^2$. Since $h$ is generally less than one, the error decreases significantly when using the two-point method. As a result, the two-point method is widely use for computing derivatives. The error can still be significant, however, if the curvature of the function changes rapidly. We will show an example of this in the results section. 

An interesting observation is that for a function such as $y = x^2$, the third-derivative, $f^{\prime\prime\prime}(x)$ is zero. Therefore the two-point scheme is exact since the error term reduces to zero. 

\subsection*{Five-Point Methods}

There exists other variants on the difference quotients for computing derivatives especially higher derivatives. We will consider one more, which is the five-point method~\cite{chapra2011numerical} for estimating the first-derivative. This is given by:
$$ f'(x) = \frac{-f(x + 2h) + 8 f(x + h) - 8 f(x - h) + f(x - 2h)}{12h} + \frac{h^4}{30} f^{(5)}(c) $$
5
Here we see that the error is of the order $h^4$. That is, if we were to set $h=0.1$, the error contribution from $h$ would be $0.00001$. 
\subsection*{Richardson Extrapolation Methods}

The previous methods use either reduced steps sizes or higher order methods to estimate a derivative. This fourth approach utilizes a completely different method, called the Richardson Extrapolation~\cite{richardson1911approximate,joyce1971survey}. This method computes two estimates for the derivative (using different $h$) from which it computes a third, more accurate estimate. 

The basic idea is that we compute the two-point method using two different values of $h$. We now have two equations with one common unknown, the derivative itself. A commonly used set of values for $h$ are $h$ and $h/2$ such that the solution for the derivative is given by:
$$ f^\prime (x) \approx \frac{4 D (h/2) - D (h)}{3} $$
where $D(h)$ is the value of the two-point estimate for the derivative when using $h$ as the step size. The Richardson Extrapolation can be iterated so as to improve the accuracy of the estimate even more. It only takes a few such iterations to obtain very accurate estimates for the derivative. In particular given that estimates can be computed at using different $h$ pairs, the results of these can also be combined. For example four initial estimates at $h, h/2, h/4, h/16$, giving three Richarsdson combinations that can be computed. From these three a further two can be computed, from which a final estimate can be computed.  These calculation form what is called an extrapolation table as shown below:

$$
\begin{array}{ll}
A_{1,1}= \mathrm{D}(h) \\[12pt]
A_{2,1}=\mathrm{D}\left(\dfrac{h}{2}\right), \hspace{1pt} A_{2,2}=\dfrac{4 A_{2,1}-A_{1,1}}{3} \\[14pt]
A_{3,1}=\mathrm{D}\left(\dfrac{h}{2^{2}}\right), \hspace{1pt} A_{3,2}=\dfrac{4 A_{3,1}-R_{2,1}}{3}, \hspace{1pt} A_{3,3}=\dfrac{16 A_{3,2}-A_{2,2}}{15} \\[14pt]
A_{4,1}=\mathrm{D}\left(\dfrac{h}{2^{3}}\right), \hspace{1pt} A_{4,2}=\dfrac{4 A_{4,1}-A_{3,1}}{3}, \hspace{1pt} A_{4,3}=\dfrac{16 A_{4,2}-A_{3,2}}{15},
\hspace{1pt} A_{4,4}=\dfrac{64 A_{4,3}-A_{3,3}}{63}  \\
\end{array}
$$

Such a table can be easily computed using software. Of interest is that the first combination of $h$ and $h/2$ ($A_{2,2}$) is equivalent to using the five-point method.

All the methods we have just described were implemented in Python 3.8 and stored in a package called SensitivityAnalysis which uses Tellurium to store models and reaction networks. The Richardson Extrapolation method was based off of an example method written in C by H. Press et al.~\cite{1992nrcabook}. In the following sections the accuracy and performance of the different numerical differentiation methods will be compared.
 
\section{Results}

\subsection{Test Cases}
\subsubsection{Mathematical Functions}
\begin{figure}
    \centering
    \includegraphics[scale=0.8]{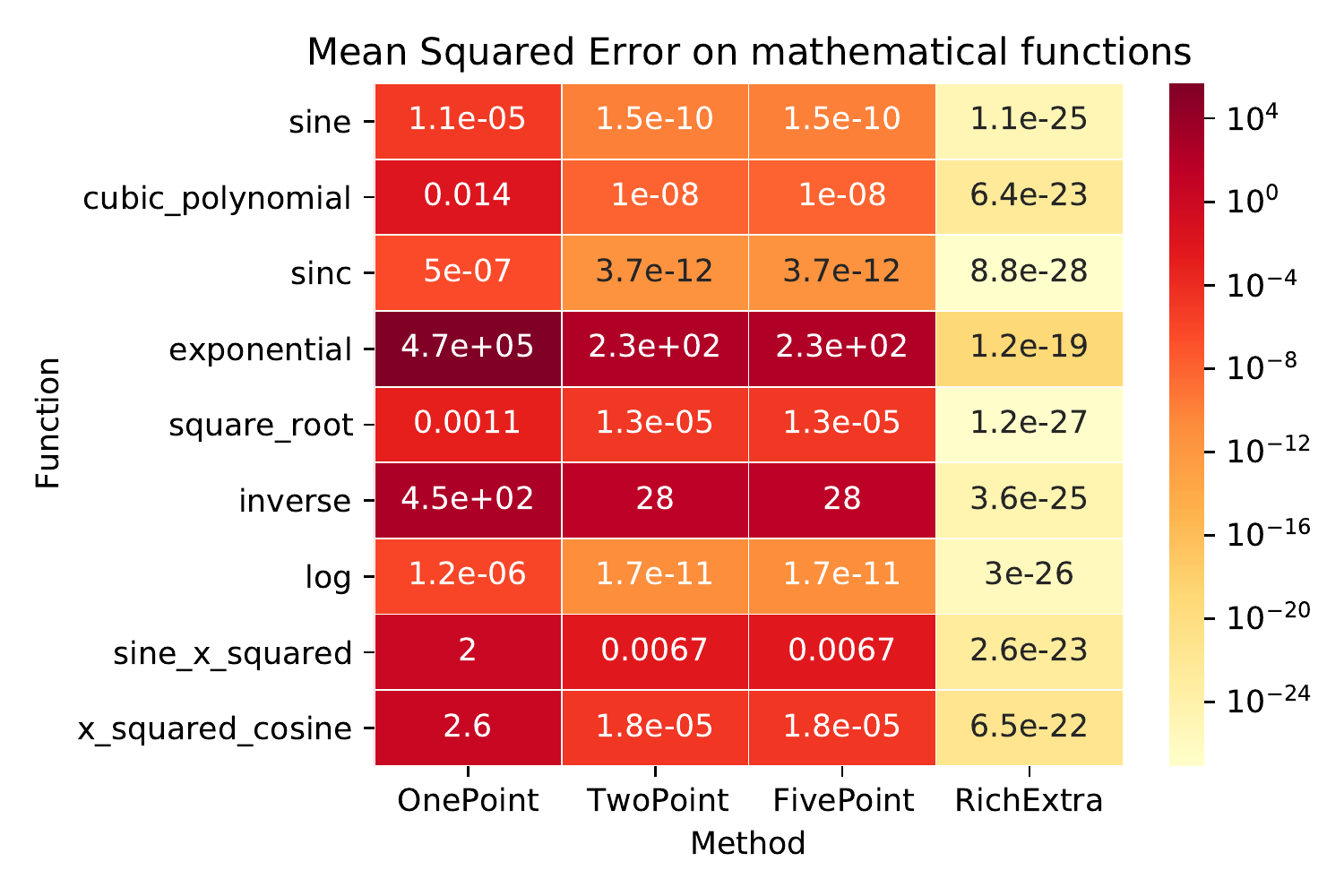}
    \caption{The mean squared errors of the computed numerical derivative via the four methods. The order of the functions in the heat map match the order shown in Table~\ref{table:test_functions}.}
    \label{fig:TestFunctionHeatMap}
\end{figure}

Mathematical functions and reaction models were used to compare the accuracy across the four different methods. Shown below in Table~\ref{table:test_functions} are the nine different mathematical test functions with their bounds~\cite{OLIVER1980145}. For each of these functions the derivative was computed across the specified range $[a, b]$ using the four different numerical differentiation methods. We would expect there to be an increase in accuracy as we move from the more simple methods to the more complex methods. This is observed in the heat map shown in Figure~\ref{fig:TestFunctionHeatMap}, where the accuracy of each method is compared using the mean squared error. For each of the methods, the estimated value of the derivative was compared to the actual value which was computed analytically through formulas shown in Table~\ref{table:test_functions}. In this plot, the darker red colors correspond to increased error while the lighter yellow colors correspond to low error. Across all the methods, there is a significant decrease in error moving from One-Point to Two-Point, Five-Point and finally Richardson Extrapolation. The highest computed error within the Richardson Methods was $1.2^{-19}$ which is still orders of magnitude below the lowest error of the other three methods.

\renewcommand{\arraystretch}{1.25}
\begin{table}[]
    \centering
\begin{tabular}{|c|c|c|c|c|}
    \hline
    No. & \(f(x)\) & $f'(x)$ & \(a\) & \(b\)\\
    \hline
    1. & $\sin{x}$ & $\cos{x}$ & $-4\pi$ & $4\pi$\\
    2. & $x^3 + 2x^2 - 3$ & $3x^2 + 4x - 3$ & -6 & 6\\
    3. & sinc $x$ & $\cos{x}/x - \sin{x}/x^2$ & $-4\pi$ & $4\pi$ \\
    4. & $\exp(x^2)$ & $2x\exp(x^2)$ & -3 & 3 \\
    5. & $\sqrt{x}$ & $1/\sqrt{x}$ & 0.05 & 1\\
    6. & $1/x$ & $-1/x^2$ & 0.05 & 1\\
    7. & $\ln{[x + \sqrt{(1+x^2)}]}$ & $1/\sqrt{x^2 + 1}$ & -2 & 5\\
    8. & $\sin{x^2}$ & $2x\cos{x^2}$ & $-4\pi$ & $4\pi$\\
    9. & $x^2\cos{2x}$ & $-2x(x\sin{2x} - \cos{2x})$ & $-4\pi$ & $4\pi$\\
    \hline
\end{tabular}
    \caption{Nine test functions evaluated over ten equally-spaced points within the interval $[a, b]$}
    \label{table:test_functions}
\end{table}

\subsubsection{Reaction Networks}

In addition to the mathematical functions, we will explore the accuracy of the numerical differentiation methods on simple reaction networks. As described earlier, reaction networks can be modeled by a series of reactions chained together. For this analysis we will use a simple three step linear pathway with two floating species and two fixed species:
$$ X_o \stackrel{v_1}{\longrightarrow} S_1 \stackrel{v_2}{\longrightarrow}
S_2 \stackrel{v_3}{\longrightarrow} X_1 $$
To compare the accuracy to a reference, the control coefficients can be computed symbolically using the following procedure. We first constructed a model using tellurium~\cite{choi2018tellurium} and then simulated it until it reached steady state. Then, we computed the elasticity coefficients using the steady-state concentrations of the species $S_1 $ and $S_2$ as well as the values of the rate constants. The general formulas used for computing the elasticities for the substrate and product are shown below~\cite{sauro2018systems}:
\begin{eqnarray*}
\epsilon^v_s &= \dfrac{k_1s}{k_1s-k_2p} &= \dfrac{v_f}{v} \\[5pt]
\epsilon^v_p &= -\dfrac{k_2p}{k_1s-k_2p} &= -\dfrac{v_r}{v} \\
\end{eqnarray*}
Using the computed elasticities ($\epsilon^{v1}_{S1}, \epsilon^{v2}_{S1}, \epsilon^{v2}_{S2}, \text{ and } \epsilon^{v3}_{S2}$) and the following formulas, we computed the control coefficients.
\begin{eqnarray*}
D &= \epsilon^2_1\epsilon^3_2 - \epsilon^1_1\epsilon^3_2 + \epsilon^1_1\epsilon^2_2 \\[10pt]
C^J_{e_1} = \frac{\epsilon^2_1\epsilon^3_2}{D} & C^J_{e_2} = -\frac{\epsilon^1_1\epsilon^3_2}{D} & C^J_{e_3} = \frac{\epsilon^1_1\epsilon^2_2}{D}
\end{eqnarray*}
\begin{align*}
C^{s_1}_{e_1} &= \frac{\epsilon^3_2 - \epsilon^2_2}{D} & C^{s_2}_{e_1} &= \frac{\epsilon^2_1}{D}\\[10pt]
C^{s_1}_{e_2} &= -\frac{\epsilon^3_2}{D} & C^{s_2}_{e_2} &= -\frac{\epsilon^1_1}{D}\\[10pt]
C^{s_1}_{e_3} &= \frac{\epsilon^2_2}{D} & C^{s_2}_{e_3} &= \frac{\epsilon^1_1-\epsilon^2_1}{D}
\end{align*}

\begin{figure}[H]
    \centering
    \includegraphics[scale=0.7]{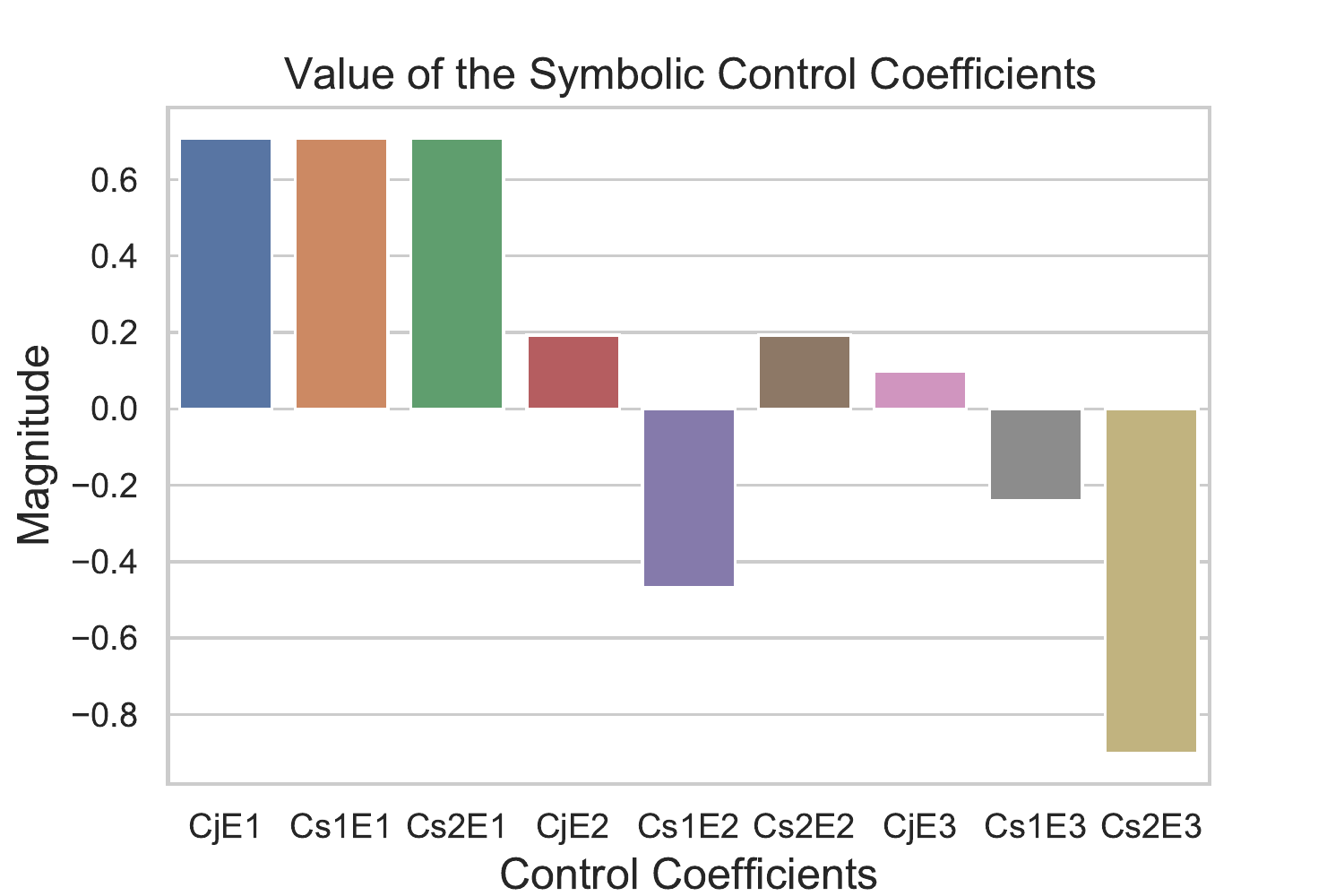}
    \caption{The computed values of the three step linear pathway Control Coefficients via analytical methods}
    \label{fig:SymbolicCCs}
\end{figure}

The values of the nine control coefficients are shown in Figure~\ref{fig:SymbolicCCs}.

\begin{figure}[H]
    \centering
    \includegraphics[scale=0.7]{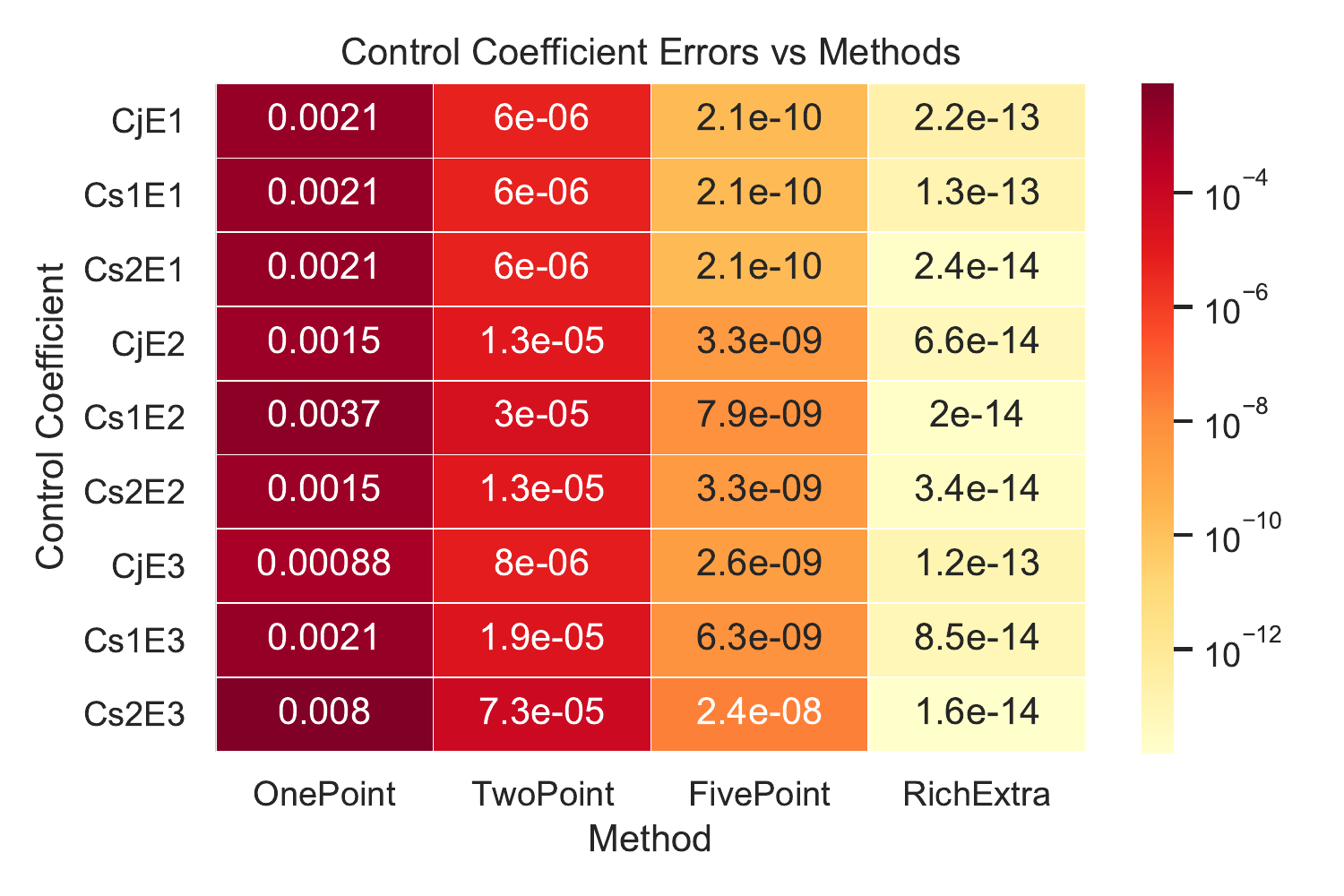}
    \caption{Shown here are the absolute errors of the computed control coefficients of the three step linear pathway. The x-axis shows the numerical differentiation type while the y-axis highlights the specific control coefficients (CjE1 represents \(C^J_{e1}\) and Cs1E1 represents \(C^{S_1}_{E_2}\)). For reference, J is the flux through the pathway, S1 represents species 1, and E1 represents the enzyme concentration of the first step}
    \label{fig:CCErrorHeatmap}
\end{figure}

\begin{figure}[H]
    \centering
    \includegraphics[scale=0.62]{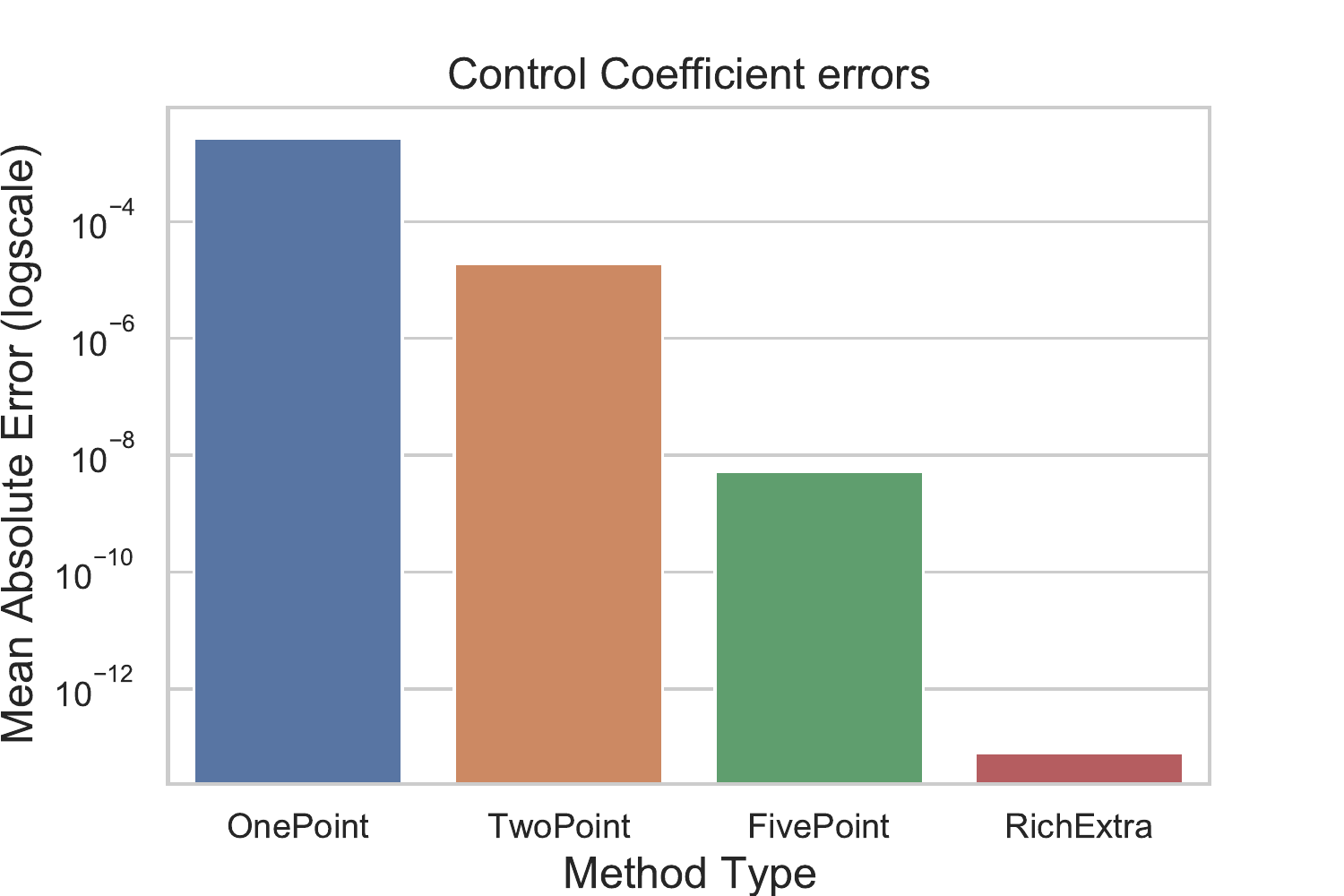}
    \caption{The mean absolute error for each numerical differentiation method computed on the three step linear pathway.}
    \label{fig:CCErrorBar}
\end{figure}

Once we had a reference for the control coefficients, the coefficients were recomputed numerically using the four methods. The Richardson Extrapolation methods performed significantly better than the other three methods with absolute errors on the order of $10^{-14}$. Looking at the heat map in Figure~\ref{fig:CCErrorHeatmap}, we can see that the absolute error decreases across individual control coefficient from One-Point to Richardson methods. This is further supported by Figure~\ref{fig:CCErrorBar}, where the errors across all nine coefficients are summarized by the mean absolute error and put on a log-scale bar chart. In this bar chart we can see that the mean absolute error appears to decrease exponentially as we move towards the more complex methods.

The results from both the mathematical test functions and reaction networks imply that the Richardson Extrapolation method is the most accurate numerical differentiation method explored here.

\subsection{Use Case: Simple Phosphorylation Cycle}
In the last section, the three step reaction network described had reasonably well behaved derivatives, characterized by moderate size and gradual changes. Thus, in the next section, s we will test the numerical differentiation techniques on a nonlinear system with derivatives of significant magnitudes. A good example of fast changing and large derivatives is found in phosphorylation cycles that exhibit ultrasensitivity~\cite{goldbeter1981amplified,ferrell2014ultrasensitivity,sauro2018systems}.

The cycle is as follows:
$$ A \stackrel{v_1}{\longrightarrow} AP$$
$$ AP \stackrel{v_2}{\longrightarrow} A$$
The rate equations used Michaelis-Menten Kinetics for both the forward and backward arms of the cycle are shown below:

$$v_1 = \frac{V_{m_1} \cdot A \cdot K}{K_{m_1} + A}$$
$$v_2 = \frac{V_{m_2}\cdot AP}{K_{m_2} + AP}$$

Using our Richardson extrapolation method we can get an accurate estimate of the Control Coefficients of concentration of AP with respect to changes in parameter K ($C^{AP}_{K}$). Shown below in Figure~\ref{fig:CycleCCs}, is the computed control coefficients over a range of different K parameter values from 0.1 to 2.0. 

\begin{figure}
    \centering
    \includegraphics[scale=0.7]{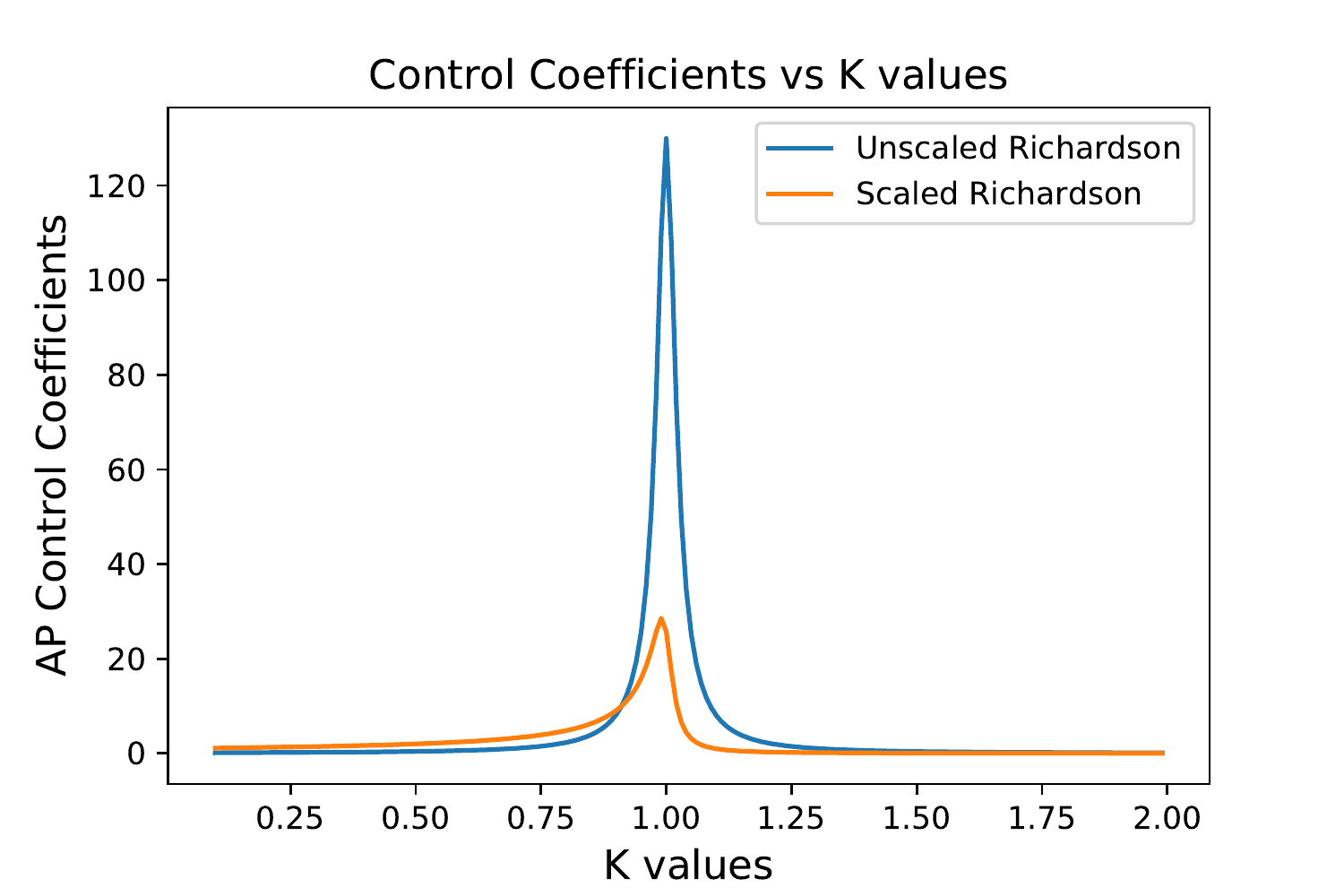}
    \caption{The computed scaled and unscaled sensitivities of concentration of AP with respect to parameter K. There is a narrow region around K=1.0 where concentration of AP is highly sensitive to changes in AP. }
    \label{fig:CycleCCs}
\end{figure}

\begin{figure}
     \centering
     \includegraphics[scale=0.7]{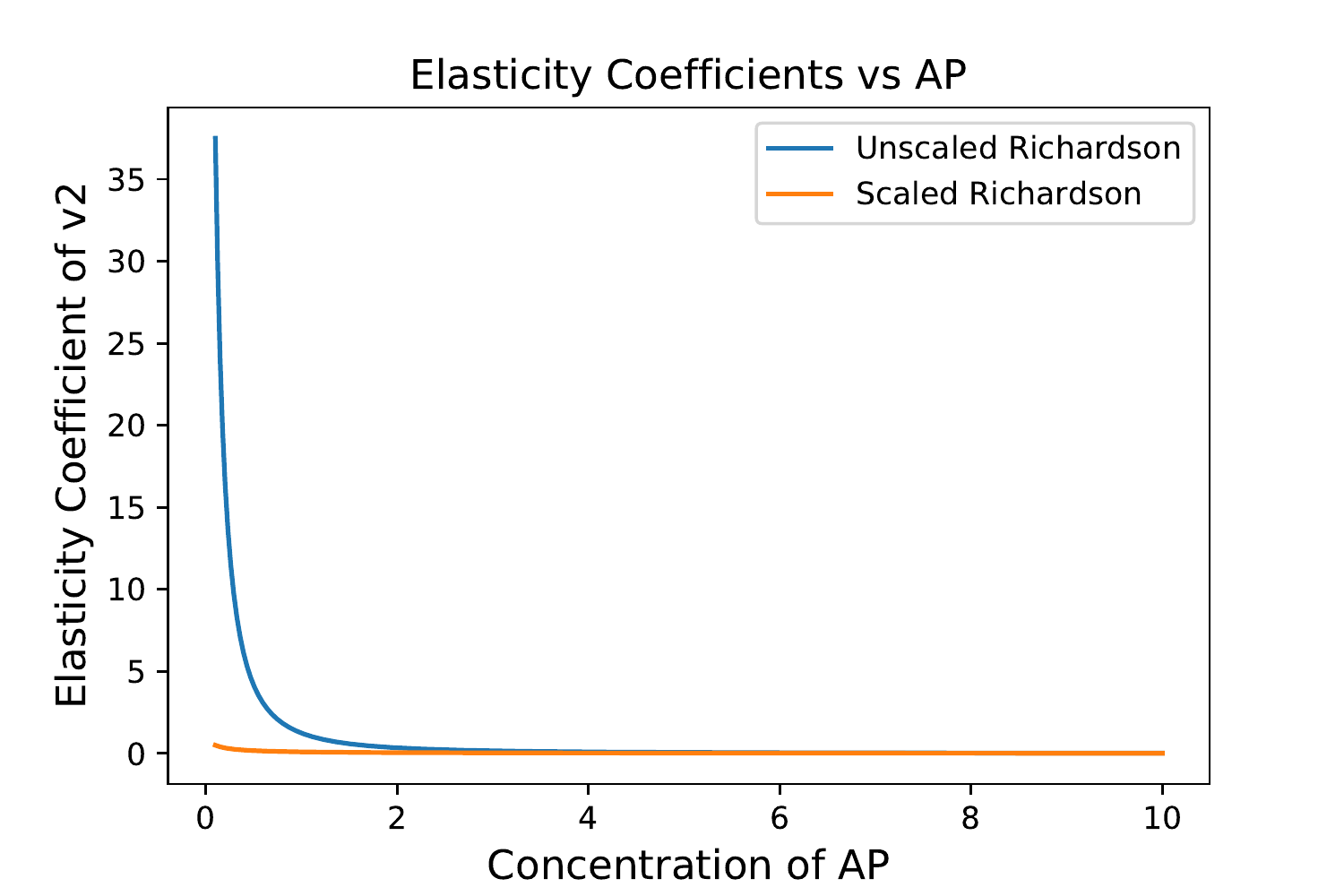}
     \caption{The computed scaled and unscaled elasticity coefficients for v2 with respect to AP. From the plot we can see that as concentration of AP increases the sensitivity of reaction v2 decreases.}
     \label{fig:CycleECs}
 \end{figure}

We can also compute the elasticity coefficients in a similar manner as shown in Figure~\ref{fig:CycleECs}. Using the Richardson Extrapolation we are able to get accurate estimates of the sensitivities of the reaction networks to changes in various constants and concentrations. After computing the coefficients, we can determine the most efficient method of perturbing the system to achieve a desired effect (e.g. increased steady-state concentration or fluxes).

\subsection{Performance}

\begin{figure}[H]
    \centering
    \includegraphics[scale=0.7]{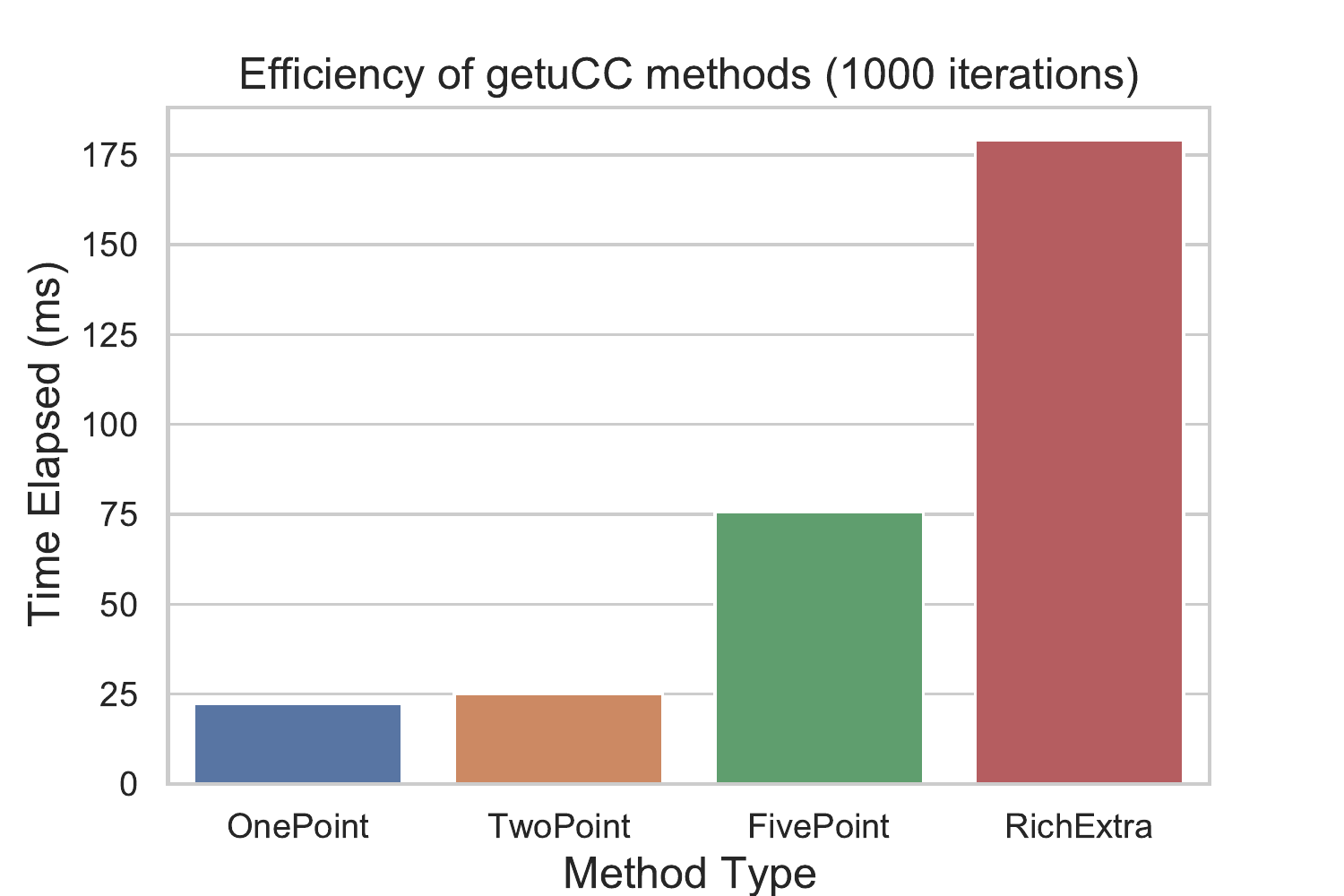}
    \caption{The time performance of the getuCC method using different numerical differentiation methods on the simple phosphorylation cycle.}
    \label{fig:getuCCtimes}
\end{figure}

After evaluating the accuracy, we now shift our focus to the performance and efficiency of the four numerical differentiation methods. In this study, we measured these metrics on the simple phosphorylation cycle. The two methods of interest are getuCC and getuEE. getuCC returns the unscaled control coefficients for a given variable (reaction or species concentration) with respect to a parameter (kinetic constant or boundary species). getuEE computes the unscaled elasticities or local sensitivities for a given reaction rate $v_i$ with respect to a given parameter. The two methods respective scaled versions are not explored here as the computational cost of scaling is trivial. Shown in Figure \ref{fig:getuCCtimes}, we can see that as we move from simple differentiation methods to more complex methods, it comes at the cost of efficiency. For $1000$ iterations of calling getuCC, the one-point method took $22.3$ ms while the Richardson Extrapolation method took $179.2$ ms. 

\begin{table}[h]
    \centering
    \begin{tabular}{|c|c|}
         \hline
         Method Type & Mean Time Elapsed (ms)\\
         \hline
         OnePoint & $22.3$\\
         TwoPoint & $25.2$\\
         FivePoint & $75.7$\\
         RichExtra & $179.2$\\
         \hline
    \end{tabular}
    \caption{Time elapsed for 1000 iterations of getuCC using specified method type on the simple phosphorylation cycle.}
    \label{tab:getuCCtimes}
\end{table}

\begin{figure}[h]
    \centering
    \includegraphics[scale=0.65]{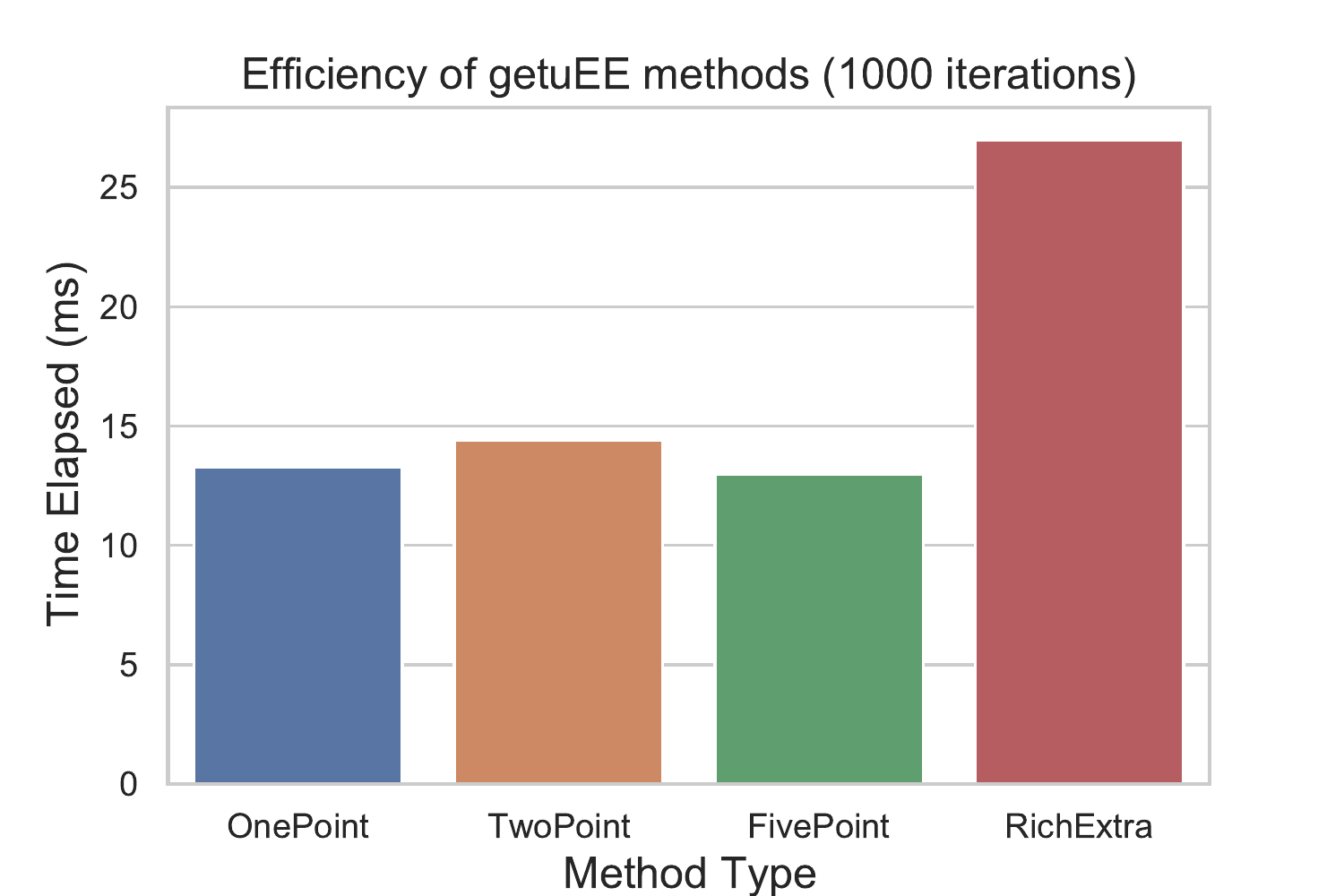}
    \caption{The time performance of the getuEE method using different numerical differentiation methods on the simple phosphorylation cycle.}
    \label{fig:getuEEtimes}
\end{figure}

In contrast, the difference in time performance of the four methods with respect to the getuEE method is much less prominent. Unlike the getuCC method, where the Richardson Extrapolation method took almost $8$ times as long, this method is only $~2$ times slower. Shown in Table \ref{tab:getuEEtimes} and Figure \ref{fig:getuEEtimes} are the elapsed times for $1000$ iterations of calling getuEE. The stark differences in times can be attributed to the necessity of computing steady state concentrations during the getuCC method.

\begin{table}[h]
    \centering
    \begin{tabular}{|c|c|}
         \hline
         Method Type & Mean Time Elapsed (ms)\\
         \hline
         OnePoint & $13.3$\\
         TwoPoint & $14.4$\\
         FivePoint & $13.0$\\
         RichExtra & $27.0$\\
         \hline
    \end{tabular}
    \caption{Time elapsed for 1000 iterations of getuEE using specified method type on the simple phosphorylation cycle.}
    \label{tab:getuEEtimes}
\end{table}

\subsection{Software Availability}

To download and use the SensitivityAnalysis Package follow the instructions listed on the {\tt README.md} file in the GitHub repository linked here: \url{https://github.com/evanyfyip/SensitivityAnalysis}

\section{Conclusion}

In this short study, we investigated a variety of methods to numerically compute steady-state sensitivities. Such sensitivities allow us to understand how much a specific enzyme contributes to the steady-state flux or species concentrations. We explored a number of different numerical methods for estimating the sensitivities. Of the methods examined (One-Point, Two-Point, Five-Point, and Richardson Extrapolation Methods), we determined that on average the Richardson Extrapolation Method performed 3 orders of magnitude better than the other methods. This was shown through comparing the mean squared errors of the computed derivative across 9 different mathematical functions. In addition, we looked at a simple biological model (a three step linear pathway) and compared the computed control coefficient errors across the methods. Again, the Richardson Extrapolation Method had significantly lower error. Thus, in terms of performance, the Richardson Extrapolation method has a much higher accuracy than other methods. This high accuracy, however, comes with a trade off in terms of time efficiency. The run times of the getuCC and getuEE methods using the different numerical differentiation techniques showed that the Richardson Extrapolation method is much slower than the other techniques. With regards to getuCC the Richardson Extrapolation Method was almost 9 times as slow as the one point numerical differentiation method. Similarly, the run time on the getuEE method was twice as slow as the one point method. Overall, the Richardson Extrapolation methods explored here provide a more accurate calculation of the control and elasticity coefficients allowing researchers to gain better insight into cellular dynamics and the potential drug targets in biological pathways. 

\printbibliography

\end{document}